\pdfoutput=1
\documentclass{article}
\usepackage{spconf,amsmath,graphicx,bbm,multirow,booktabs}
\usepackage[numbers,sort&compress]{natbib}

\title{Unrestricted Global Phase Bias-Aware Single-channel Speech Enhancement with Conformer-based Metric GAN}
\name{Shiqi Zhang$^*$, Zheng Qiu$^*$, Daiki Takeuchi$^\dag$, Noboru Harada$^\dag$, Shoji Makino$^*$\thanks{This work was supported by JSPS KAKENHI Grant Number 23H03423.}}
\address{$^*$ Waseda University, Japan~~and~~$^\dag$ NTT Coporation, Japan}
\begin{document}
\ninept
\maketitle
\begin{abstract}
With the rapid development of neural networks in recent years, the ability of various networks to enhance the magnitude spectrum of noisy speech in the single-channel speech enhancement domain has become exceptionally outstanding.
However, enhancing the phase spectrum using neural networks is often ineffective, which remains a challenging problem.
In this paper, we found that the human ear cannot sensitively perceive the difference between a precise phase spectrum and a biased phase (BP) spectrum.
Therefore, we propose an optimization method of phase reconstruction, allowing freedom on the global-phase bias instead of reconstructing the precise phase spectrum. We applied it to a Conformer-based Metric Generative Adversarial Networks (CMGAN) baseline model, which relaxes the existing constraints of precise phase and gives the neural network a broader learning space.
Results show that this method achieves a new state-of-the-art performance without incurring additional computational overhead.
\end{abstract}
\section{Introduction}
\label{sec:intro}

Speech enhancement~\cite{speechenhancement} is a technique that involves processing noisy speech signals to output relatively clean speech signals. It has a wide range of applications, spanning communication devices, intelligent interactive devices, hearing aids, and more. A good speech enhancement front-end technology can not only provide a better communication experience but also lead to improved accuracy in automatic speech recognition (ASR)~\cite{asr, se2asr}.

\vspace{-2.mm}
\begin{figure}[htp]
\begin{minipage}[b]{1.0\linewidth}
  \centering
  \centerline{\includegraphics[width=8.5cm]{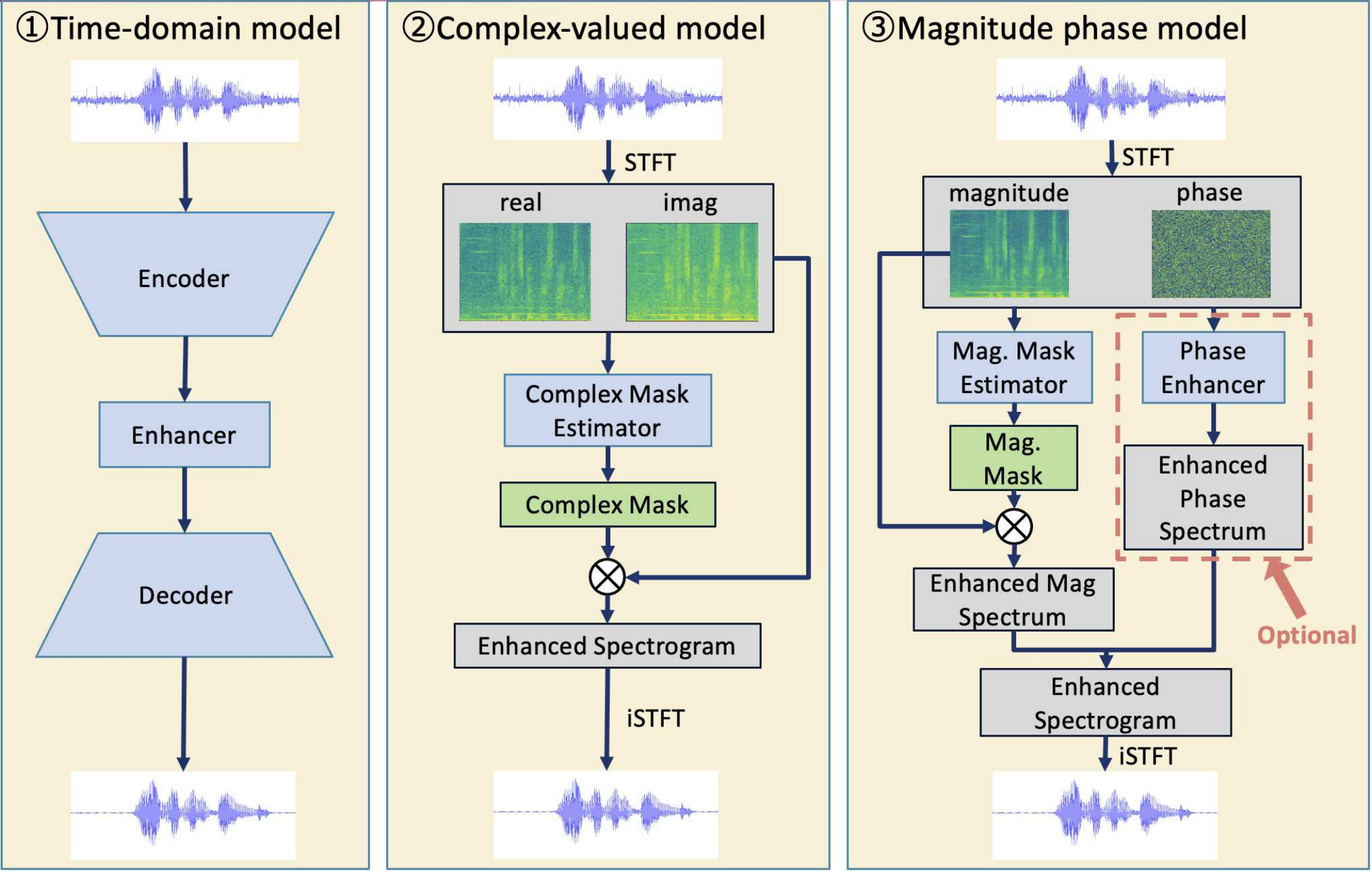}}
\end{minipage}
\vspace{-4.mm}
\caption{Three different types of speech enhancement models}
\label{fig:three_type_models}
\end{figure}
\vspace{-2.mm}

With the rapid development of neural networks in recent years, a large number of excellent single-channel speech enhancement neural network models have emerged in the field of acoustic signal processing, including the classic magnitude-phase models~\cite{metricgan, metricgan+, db-aiat, mpsenet, phasen, mask, tfmaskgan} and the recently popular complex-valued models~\cite{cirmc, cirmj, pfpl, dccrn, dpt-fsnet, tridnetse} and time-domain models~\cite{segan, seconformer, tasnet, convtasnet} as shown in Fig. \ref{fig:three_type_models}. Among these, the magnitude-phase models and complex-valued models both process noisy speech signals in the time-frequency domain. In this processing flow, the noisy speech is converted to the time-frequency domain using the short-time Fourier transform (STFT), the spectrogram is enhanced, and the enhanced spectrogram is converted back to the time-domain signal and output using the inverse STFT (iSTFT). The difference between these two models is that the former estimates a real-valued mask matrix applied to the magnitude and a cleaner phase spectrum (optional), while the latter estimates a complex-valued mask~\cite{cirmc, cirmj} matrix applied to the entire complex spectrum. The time-domain model usually consists of an encoder, a decoder, and a network in the middle for enhancement. After inputting the signal, it is directly encoded in the time domain, and then decoded after enhancement in the encoding domain to obtain a cleaner audio signal. Although the implementation details of the above three methods are different, they all aim at reconstructing precise speech signals that have accurate magnitude and phase.

However, due to the complexity of the phase, accurately reconstructing the phase spectrum~\cite{griffinlim, phasen, phaserecon} has posed significant challenges for existing neural networks~\cite{kim2022phase, masuyama2019deep, thien2023inter, thien2021two, masuyama2020deep}. Through our own experiments, we discovered that it is difficult for the human ear to distinguish between an accurate phase and a globally biased phase spectrum~\cite{sens, pulse}. Utilizing this characteristic, we propose a new optimization method based on unrestricted globally biased phase reconstruction that improves the performance of speech enhancement without increasing the number of model parameters or the computational cost, and achieves new state-of-the-art (SoTA) results.

\vspace{-3mm}
\section{Biased phase perceptual experiment}
\vspace{-3mm}

\label{sec:hear_test}

The experiment conducted in this section reveals that the human auditory system is not sensitive to precise phase. A perfect reconstruction of the phase spectrum is not necessary, and global phase bias can therefore be ignored. This provides one more dimension to the solution space that does not affect the perceptual quality, thereby facilitating easier convergence of neural networks.

\begin{figure}[htp]
\begin{minipage}[b]{1.0\linewidth}
  \centering
  \centerline{\includegraphics[height=6.cm, width=8.5cm]{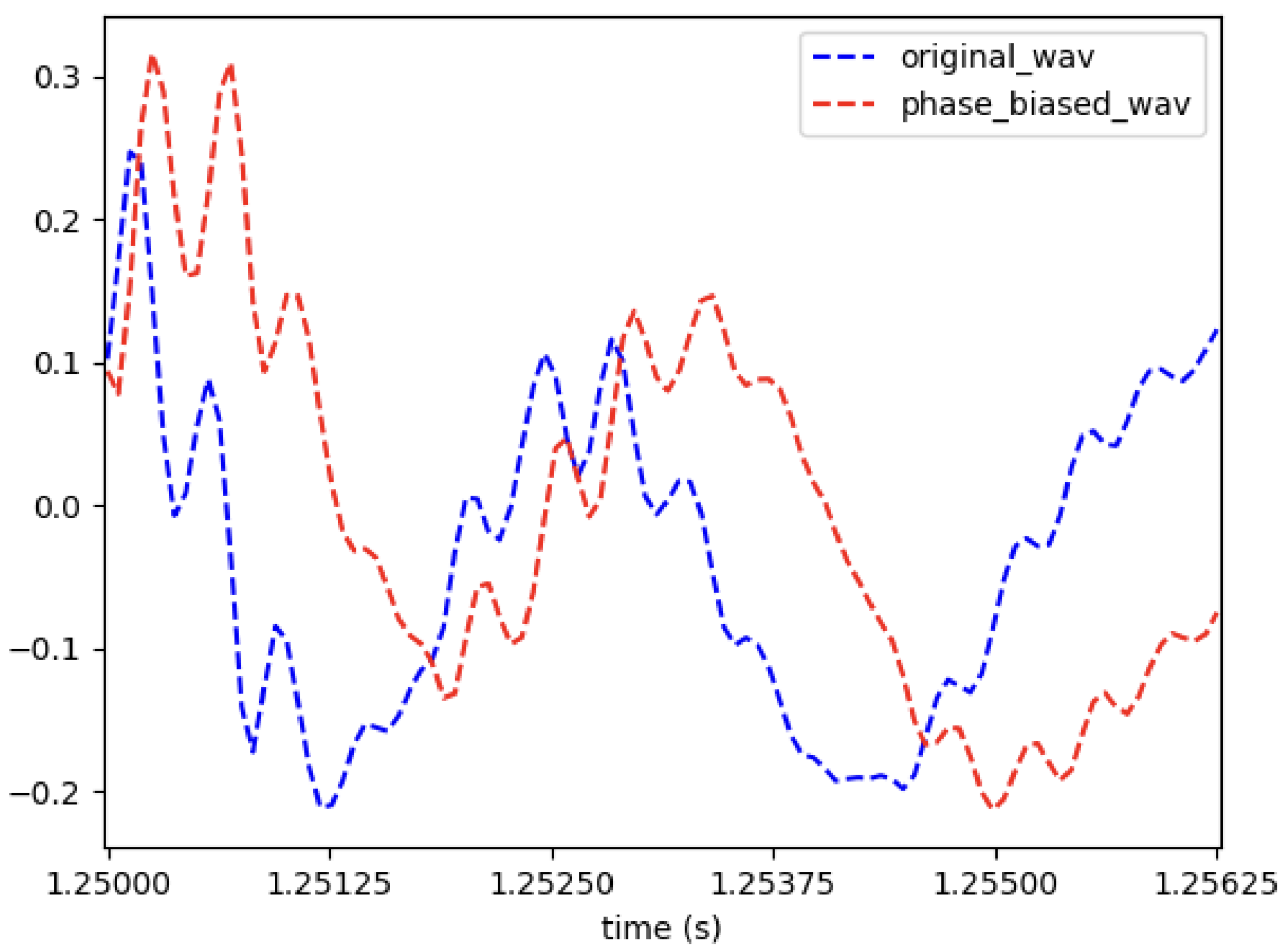}}
\end{minipage}
\vspace{-6mm}
\caption{Comparison of original phase reconstructed signal and biased phase reconstructed signal (partial)}
\label{img:wavs}
\end{figure}

First, we processed 100 audio samples using STFT with a Hamming window $h$ and a shift value $\theta$ as shown in (\ref{eq:biased_stft_1}). The modification only involves adding a random global angle value, $\theta\in[-\pi,\pi)$, to the phase part of the original spectrum without affecting the magnitude spectrum at all. Therefore, (\ref{eq:biased_stft_1}) merely produces a time-frequency spectrogram with a global phase bias. Subsequently, the two types of spectrograms obtained are ultilized for time-domain signal reconstruction using iSTFT. \big(Note that (\ref{eq:biased_stft_1}) can also be written in the form of (\ref{eq:biased_stft_2}) if we want to show that this processing is not a simple time shift for the waveform. For different frequency components $\omega$, the time shift amount $\frac{\theta}{\omega}$ is also different. The reconstructed signal is inconsistent with the original signal in terms of the time-domain waveform, as shown in Fig. \ref{img:wavs}.\big)

\vspace{-3mm}
\begin{gather}
    X(t,\omega)=\int^{\infty}_{-\infty}x(\tau)h(\tau-t)e^{-j(\omega\tau+\theta)}d\tau
    \label{eq:biased_stft_1}\\
    X(t,\omega)=\int^{\infty}_{-\infty}x(\tau)h(\tau-t)e^{-j\omega(\tau+\frac{\theta}{\omega})}d\tau
    \label{eq:biased_stft_2}
\end{gather}
\vspace{-5mm}

\vspace{-3mm}
Next, we conducted a hearing test with four listeners. They first listened to the original audio and then to the signals reconstructed using iSTFT after transformation with STFT and (\ref{eq:biased_stft_1}), and judged and recorded which of the two latter signals was closer to the original audio. If the judgment was correct, it was counted as 1 for ``True''; otherwise, it was counted as 1 for ``False''. The results, as listed in Table \ref{tab:hear_test}, indicate that the ratio of True to False counts is close to 50\%. This indicates a similar phenomenon to~\cite{sens, pulse}, namely, that the human ear has little ability to accurately distinguish between the phase spectrum and the phase spectrum with a global bias.

\vspace{-5mm}
\begin{table}[ht]
\caption{Hearing results of phase unbiased and biased signals}
\label{tab:hear_test}
\setlength{\tabcolsep}{3.8mm}{
\begin{tabular}{c|cccc|c}
\toprule
Result & $P_1$ & $P_2$ & $P_3$ & $P_4$ & Total \\ \hline\hline
True   & 46    & 58    & 55    & 50    & 209 (52.25\%) \\ \hline
False  & 54    & 42    & 45    & 50    & 191 (47.75\%) \\ \bottomrule
\end{tabular}}
\end{table}
\vspace{-2mm}


We also calculated the average Perceptual Evaluation of Speech Quality (PESQ)~\cite{pesq}, Segment Signal-to-Noise Ratio (SegSNR)~\cite{segsnr}, and Scale-invariant Source-to-Noise Ratio (SiSNR)~\cite{tasnet} for the signals reconstructed using STFT-iSTFT and those reconstructed using (\ref{eq:biased_stft_1})-iSTFT. (SegSNR and SiSNR are limited to the range of [-10.000, 35.000] dB.) Table \ref{tab:pesq_ssnr_sisnr} shows results consistent with the perception experiment in Table \ref{tab:hear_test} and Figure \ref{img:wavs}. PESQ is mostly unaffected by the biased phase, whereas the time-domain metrics SegSNR and SiSNR are significantly impacted. These findings also indicate that time-domain metrics in the field of signal processing do not directly represent the perception effects of audio.

\vspace{-4mm}
\begin{table}[htb]
\caption{Metrics on phase unbiased and biased signals}
\label{tab:pesq_ssnr_sisnr}
\setlength{\tabcolsep}{5.8mm}{
\begin{tabular}{c|c|c|c}
\toprule
Signal                  & PESQ & SegSNR & SiSNR \\ \hline \hline
Unbiased                & 4.644    & 35.000      & 35.000     \\ \hline
Biased                  & 4.636    & 0.561      & 1.627     \\ \bottomrule
\end{tabular}}
\end{table}

\section{Methodology}
\label{sec:method}
\vspace{-2mm}
In this paper, we selected CMGAN~\cite{cmgan} as the baseline and improved the optimization method using the phenomenon discussed in Sec. \ref{sec:hear_test}.
\vspace{-2mm}

\subsection{Baseline: CMGAN}

CMGAN is a generalized magnitude-phase model that generates a real-valued mask matrix and a complex-valued residual matrix through a stacked network structure. The mask is multiplied by the original magnitude spectrum to obtain a relatively clean magnitude spectrum, which is then reconstructed into a spectrogram together with the original noisy phase spectrum. The complex-valued residual matrix is then added to this spectrogram, achieving the effect of simultaneously estimating the magnitude and phase. The original optimization function of CMGAN ($\mathcal{L}_{ori}$) is a composite loss function (\ref{eq:ori_loss}) consisting of the following weighted ($\lambda_{1\sim4}$) components: magnitude Mean Square Error (MSE) Loss ($\mathcal{L}_{mag}$), spectrum real and imaginary part MSE Loss ($\mathcal{L}_{ri}$), time-domain L1 Loss ($\mathcal{L}_{time}$), and Adversarial Loss ($\mathcal{L}_{adv}$). These are jointly used for optimizing the generator network.

\vspace{-3mm}
\begin{equation}
    \mathcal{L}_{ori} = \lambda_{1}\mathcal{L}_{mag}+\lambda_{2}\mathcal{L}_{ri}+\lambda_{3}\mathcal{L}_{time}+\lambda_{4}\mathcal{L}_{adv}
    \label{eq:ori_loss}
\end{equation}

As for the discriminator part, it can be viewed as a normalized PESQ predictor. It undergoes joint training consisting of two loss components. The first is to narrow the gap between the real normalized PESQ and the discriminator's prediction for the output of the generator and the clean signal. The second is to make the predictions for the same clean signal closer to 1, ensuring accuracy in assessing signal quality.
\vspace{-3mm}

\subsection{Proposed methods}

As shown in Sec. \ref{sec:hear_test}, a precise phase is unnecessary for human ear perception. Therefore, we propose the following methods, hoping to ignore any global bias on the phase spectrum, relax the existing constraints of the precise phase, and give the neural network a broader learning space, thereby more easily achieving better enhancement effects.
\vspace{-2.5mm}

\subsubsection{Unrestricted phase bias (UPB)-aware loss function}
\label{sssc:bpalf}

To estimate a globally biased phase spectrum, we naturally thought of using the MSE loss of its partial derivatives for model optimization, due to the property that the derivative of a constant is zero. At the same time, to avoid the impact of meaningless accurately estimating the phase, we removed the $\mathcal{L}_{ri}$ and $\mathcal{L}_{time}$ parts from the original CMGAN optimization function.

Specifically, we first use (\ref{eq:tpd}) and (\ref{eq:fpd}) to compute derivatives of the phase spectrum's time and frequency, referred to as Time Phase Derivative (TPD~\cite{masu}) $\delta_{\tau}\in\mathbbm{R}^{(T-1)\times F}$ and Frequency Phase Derivative (FPD~\cite{masu}) $\delta_{\omega}\in\mathbbm{R}^{T\times (F-1)}$, respectively.
\vspace{-1mm}
\begin{gather}
    \boldsymbol{\delta}_{\tau}(t,f)=\boldsymbol{\phi}(t+1,f)-\boldsymbol{\phi}(t,f)
    \label{eq:tpd}\\
    \boldsymbol{\delta}_{\omega}(t,f)=\boldsymbol{\phi}(t,f+1)-\boldsymbol{\phi}(t,f)
    \label{eq:fpd}
\end{gather}
\vspace{-3.5mm}

Here, $t$ and $f$ represent the time and frequency coordinates of the elements in the spectrum matrix, $\boldsymbol{\phi}\in\mathbbm{R}^{T\times F}$ represents the phase angle, $T$ is the total number of time frames, and $F$ is the total number of frequency bins.
However, this method has a wrapping problem during the calculation process. Specifically, when the absolute sum of two phase angles exceeds $\pi$, for example, $\theta_{1}=\frac{3}{4}\pi$, $\theta_{2}=-\frac{3}{4}\pi$, their algebraic difference is $\theta_{2}-\theta_{1}=-\frac{3}{2}\pi$. However, $\theta_{1}+\frac{1}{2}\pi=\frac{5}{4}\pi=-\frac{3}{4}\pi+2\pi$ and $\theta_{2}=-\frac{3}{4}\pi$ are equivalent, i.e., increasing $\theta_{1}$ by $\frac{1}{2}\pi$ results in $\theta_{2}$.

Therefore, we use a common method (\ref{eq:ang_sub}) similar to the one in~\cite{ifd} to calculate the angle difference and solve the wrapping issue.
\begin{equation}
    \mathcal{S}(\theta_{1},\theta_{2})=arctan\big(tan(\theta_{1}-\theta_{2})\big)
    \label{eq:ang_sub}
\end{equation}
As a result, the wrapped TPD $\delta_{w\tau}\in\mathbbm{R}^{(T-1)\times F}$ and FPD $\delta_{w\omega}\in\mathbbm{R}^{T\times (F-1)}$ can be represented by (\ref{eq:utpd}) and (\ref{eq:ufpd}), respectively.
\begin{gather}
    \boldsymbol{\delta}_{w\tau}(t,f)=\mathcal{S}\big(\boldsymbol{\phi}(t+1,f),\boldsymbol{\phi}(t,f)\big)
    \label{eq:utpd}\\
    \boldsymbol{\delta}_{w\omega}(t,f)=\mathcal{S}\big(\boldsymbol{\phi}(t,f+1),\boldsymbol{\phi}(t,f)\big)
    \label{eq:ufpd}
\end{gather}
Hence, the loss function $\mathcal{L}_{upb}$ for estimating unrestricted biased phase can be calculated using (\ref{eq:bp_loss}).
\begin{equation}
    \begin{aligned}
        \mathcal{L}_{upb}&=\frac{1}{2}\mathbbm{E}_{\boldsymbol{\delta}_{w\tau}(t,f),\hat{\boldsymbol{\delta}}_{w\tau}(t,f)}[\Vert\mathcal{S}\big(\boldsymbol{\delta}_{w\tau}(t,f),\hat{\boldsymbol{\delta}}_{w\tau}(t,f)\big)\Vert^{2}]\\&+\frac{1}{2}\mathbbm{E}_{\boldsymbol{\delta}_{w\omega}(t,f),\hat{\boldsymbol{\delta}}_{w\omega}(t,f)}[\Vert\mathcal{S}\big(\boldsymbol{\delta}_{w\omega}(t,f),\hat{\boldsymbol{\delta}}_{w\omega}(t,f)\big)\Vert^{2}]
    \end{aligned}
    \label{eq:bp_loss}
\end{equation}
Finally, the model loss function of CMGAN is modified here to (\ref{eq:bp_whole_loss}).
\begin{equation}
    \mathcal{L}_{1}=\lambda_{1}\mathcal{L}_{mag}+\lambda_{4}\mathcal{L}_{adv}+\lambda_{5}\mathcal{L}_{upb}
    \label{eq:bp_whole_loss}
\end{equation}
The loss factors $\lambda_{2}{L}_{ri}$ and $\lambda_{3}{L}_{time}$ have been removed from (\ref{eq:ori_loss}). This method directly unrestricts the phase bias instead of reconstructing the phase spectrum from TPD and FPD through an additional model like~\cite{recurrent}.

\vspace{-2mm}
\subsubsection{Magnitude-based Weighted UPB-aware loss function}
\label{sssc:awbpalf}
In Sec. \ref{sssc:bpalf}, we did not weight the individual time-frequency bins in the TPD and FPD spectra, which implies that all TPD and FPD were considered to have the same perceptual impact on the final enhanced signal. However, this contradicts common sense. Consider the following scenario: there are two time-frequency bins, and if the magnitude of one is very small (i.e., close to 0) while the magnitude of the other is very large, will the corresponding TPD and FPD have the same perceptual impact? Clearly, they will not, because when converting back to the time domain, the time-frequency bin with a larger magnitude will produce stronger sinusoidal signals, whereas the signals from the time-frequency bin close to 0 can be almost ignored. At this point, its phase offset value can also be ignored.

On the basis of the above insight, we propose the magnitude-based weighted UPB loss function on the basis of (\ref{eq:bp_whole_loss}). First, we use (\ref{eq:cmp_amp}) to obtain the power-compressed magnitude spectrum $\mathbf{M}_{cmp}(t,f)\in\mathbbm{R}^{T\times F}$.
\begin{equation}
    \mathbf{M}_{cmp}(t,f)=\big(\mathbf{M}(t,f)\big)^{c}=\big(|\mathbf{X}(t,f)|\big)^{c}
    \label{eq:cmp_amp}
\end{equation}
Here, $\mathbf{X}$ represents the spectrogram of the audio processed by STFT.

After obtaining $\mathbf{M}_{cmp}$, we use (\ref{eq:wutpd}) and (\ref{eq:wufpd}) to compute the magnitude-weighted wrapped TPD $\delta_{ww\tau}\in\mathbbm{R}^{(T-1)\times F}$ and FPD $\delta_{ww\omega}\in\mathbbm{R}^{T\times (F-1)}$. $\Sigma$ here means to sum up.
\begin{gather}
    \boldsymbol{\delta}_{ww\tau}(t,f)=\frac{\mathbf{M}_{cmp}(t+1,f)+\mathbf{M}_{cmp}(t,f)}{\Sigma\big(\mathbf{M}_{cmp}(t+1,f)+\mathbf{M}_{cmp}(t,f)\big)}\boldsymbol{\delta}_{w\tau}(t,f)
    \label{eq:wutpd}\\
    \boldsymbol{\delta}_{ww\omega}(t,f)=\frac{\mathbf{M}_{cmp}(t,f+1)+\mathbf{M}_{cmp}(t,f)}{\Sigma\big(\mathbf{M}_{cmp}(t,f+1)+\mathbf{M}_{cmp}(t,f)\big)}\boldsymbol{\delta}_{w\omega}(t,f)
    \label{eq:wufpd}
\end{gather}
Consequently, the magnitude-weighted UPB loss $\mathcal{L}_{wupb}$ can be expressed as (\ref{eq:awbp_loss}).
\begin{equation}
    \begin{aligned}
        \mathcal{L}_{wupb}&=\frac{1}{2}\mathbbm{E}_{\boldsymbol{\delta}_{wu\tau}(t,f),\hat{\boldsymbol{\delta}}_{wu\tau}(t,f)}[\Vert\mathcal{S}\big(\boldsymbol{\delta}_{wu\tau}(t,f),\hat{\boldsymbol{\delta}}_{wu\tau}(t,f)\big)\Vert^{2}]\\&+\frac{1}{2}\mathbbm{E}_{\boldsymbol{\delta}_{wu\omega}(t,f),\hat{\boldsymbol{\delta}}_{wu\omega}(t,f)}[\Vert\mathcal{S}\big(\boldsymbol{\delta}_{wu\omega}(t,f),\hat{\boldsymbol{\delta}}_{wu\omega}(t,f)\big)\Vert^{2}]
    \end{aligned}
    \label{eq:awbp_loss}
\end{equation}
It is important to note that this equation is the loss function during training optimization. The weights should be calculated using the clean magnitude as the standard, so even in $\hat{\boldsymbol{\delta}}_{ww\tau}(t,f)$ and $\hat{\boldsymbol{\delta}}_{ww\omega}(t,f)$, the weighted portion still uses the clean audio magnitude spectrum $\mathbf{M}_{cmp}$ instead of the estimated magnitude spectrum $\hat{\mathbf{M}}_{cmp}$.

Finally, the overall loss function $\mathcal{L}_{2}$ of CMGAN optimized by the magnitude-weighted biased phase loss function can be expressed as (\ref{eq:awbp_whole_loss}).
\begin{equation}
    \mathcal{L}_{2}=\lambda_{1}\mathcal{L}_{mag}+\lambda_{4}\mathcal{L}_{adv}+\lambda_{6}\mathcal{L}_{wupb}
    \label{eq:awbp_whole_loss}
\end{equation}

\vspace{-5mm}
\subsubsection{UPB-aware discriminator}
\label{sssc:bpad}
We have also incorporated the aforementioned properties of the biased phase into the Discriminator of CMGAN. The original discriminator input consists only of the magnitude spectra of clean and estimated speech, ($\mathbf{M}, \hat{\mathbf{M}}$), to which we add four additional inputs: wrapped TPD and FPD of clean and estimated speech ($\boldsymbol{\delta}_{w\tau}, \boldsymbol{\delta}_{w\omega}, \hat{\boldsymbol{\delta}}_{w\tau}, \hat{\boldsymbol{\delta}}_{w\omega}$).

Initially, we concatenate the above spectra together, using (\ref{eq:input_disc}) to generate the corresponding input matrices $\mathbf{A}$ and $\hat{\mathbf{A}}$ for the discriminator. Here, $\mathbf{A}\in\mathbbm{R}^{T\times F}$ represents the input matrix corresponding to the clean speech signal, and $\hat{\mathbf{A}}\in\mathbbm{R}^{T\times F}$ corresponds to the estimated signal.

\vspace{-3mm}
\begin{equation}
    \mathbf{A}=[\boldsymbol{\delta}_{w\tau}:\boldsymbol{\delta}_{w\omega}:\mathbf{M}]
    \label{eq:input_disc}
\end{equation}

\vspace{-1mm}
In this formula, `:' represents the concatenation operation. Since the shapes of $\delta_{w\tau}\in\mathbbm{R}^{(T-1)\times F}$, $\delta_{w\omega}\in\mathbbm{R}^{T\times (F-1)}$, and $\mathbf{M}\in\mathbbm{R}^{T\times F}$ do not match, we first need to pre-perform a one-length zero padding on the boundaries of $\delta_{w\tau}$ and $\delta_{w\omega}$ to make the shapes consistent before performing the concatenation operation.

Then, the prediction process of the normalized PESQ score by the discriminator $\mathcal{D}(\cdot)$, $Q_{PESQ}=\frac{PESQ-1}{3.65}$, can be represented as in (\ref{eq:pesq_score}).
\begin{equation}
    \hat{Q}_{PESQ}=\mathcal{D}(\mathbf{A},\hat{\mathbf{A}})
    \label{eq:pesq_score}
\end{equation}
Here, $\hat{Q}_{PESQ}$ represents the predicted value of the normalized PESQ score from the discriminator.
The optimization function of the discriminator is shown in (\ref{eq:disc_loss}).
\begin{equation}
    \begin{aligned}   
    \mathcal{L}_{D} &= \mathbbm{E}_{\mathbf{A},\mathbf{A}}\big[\Vert\mathcal{D}(\mathbf{A},\mathbf{A}) -1\Vert^{2}\big]\\
     &+ \mathbbm{E}_{\mathbf{A},\hat{\mathbf{A}}}\big[\Vert\mathcal{D}(\mathbf{A},\hat{\mathbf{A}})-Q_{PESQ}\Vert^{2}\big]
    \end{aligned}
    \label{eq:disc_loss}
\end{equation}

\vspace{-1mm}
At this point, the new adversarial loss of the generator based on the UPB discriminator, $\mathcal{L}_{upb-adv}$, is given by (\ref{eq:bp_adv_loss}).

\vspace{-1mm}
\begin{equation}
    \mathcal{L}_{upb-adv} = \mathbbm{E}_{\mathbf{A},\hat{\mathbf{A}}}\big[\Vert\mathcal{D}(\mathbf{A},\hat{\mathbf{A}})-1\Vert^{2}\big]
    \label{eq:bp_adv_loss}
\end{equation}

Ultimately, the overall loss function of the optimized CMGAN, $\mathcal{L}_{3}$, can be expressed as in (\ref{eq:disc_whole_loss}).
\begin{equation}
    \mathcal{L}_{3}=\lambda_{1}\mathcal{L}_{mag}+\lambda_{6}\mathcal{L}_{wupb}+\lambda_{7}\mathcal{L}_{upb-adv}
    \label{eq:disc_whole_loss}
\end{equation}

\vspace{-4mm}
\begin{table*}[t!]
\caption{Performance comparison on the Voice Bank+DEMAND dataset. ``/" denotes that the current model does not have this hyper-parameter. ``-" denotes that the result is not provided. (`T' stands for True, `F' stands for False.)}
\label{tab:result}
\setlength{\tabcolsep}{1.43mm}{
\begin{tabular}{c|c|cccc|c|cccc}
\toprule
\multirow{2}{*}{Model}
& \multirow{2}{*}{Input type}
& \multicolumn{4}{c|}{UPB}
& \multirow{2}{*}{No. of paras. (M)}
& \multirow{2}{*}{PESQ}
& \multirow{2}{*}{CSIG}
& \multirow{2}{*}{COVL}
& \multirow{2}{*}{STOI}
\\ \cline{3-6}
& & Loss & Weight & Disc. & Data aug. & & & & & \\ \cline{1-1} \cline{6-11}\hline \hline
Noisy
& / & / & / & / & / & -- & 1.97 & 3.35 & 2.63 & 0.91 \\ \hline \hline
PHASEN~\cite{phasen}
& Magnitude + Phase & / & / & / & / & -- & 2.99 & 4.21 & 3.62 & -- \\ \hline
MetricGAN+~\cite{metricgan+}
& Magnitude & / & / & / & / & -- & 3.15 & 4.14 & 3.64 & -- \\ \hline
MANNER~\cite{manner}
& Time domain & / & / & / & / & -- & 3.21 & 4.53 & 3.91 & 0.95 \\ \hline
D$^{2}$Net~\cite{d2net}
& Complex & / & / & / & / &  -- & 3.27 & 4.63 & 3.92 & \textbf{0.96} \\ \hline
DPT-FSNet~\cite{dpt-fsnet}
& Complex & / & / & / & / & 0.91 & 3.33 & 4.58 & 4.00 & \textbf{0.96} \\ \hline
D2Former~\cite{d2former}
& Complex & / & / & / & / & 0.87 & 3.43 & 4.66 & 4.22 & \textbf{0.96} \\ \hline
MP-SENet~\cite{mpsenet}
& Magnitude + Phase & / & / & / & / & 2.05 & 3.50 & 4.73 & 4.22 & \textbf{0.96} \\ \hline 
CMGAN (baseline)~\cite{cmgan}
& Magnitude + Complex & / & / & / & / & 1.83 & 3.41 & 4.63 & 4.12 & \textbf{0.96} \\ \hline \hline
\multirow{4}{*}{\begin{tabular}[c]{@{}c@{}}\textbf{UPB-CMGAN} \\ (\textbf{proposed})\end{tabular}}
& Magnitude + Complex & T & F & F & F & 1.83 & 3.46 & 4.73 & 4.21 & \textbf{0.96} \\ \cline{2-11} 
& Magnitude + Complex & T & T & F & F & 1.83 & 3.49 & 4.75 & 4.25 & \textbf{0.96} \\ \cline{2-11} 
& Magnitude + Complex & T & T & T & F & 1.83 & 3.52 & 4.75 & 4.24 & \textbf{0.96} \\ \cline{2-11} 
& Magnitude + Complex & T & T & T & T & 1.83 & \textbf{3.55} & \textbf{4.78} & \textbf{4.28} & \textbf{0.96} \\ \bottomrule
\end{tabular}}
\end{table*}

\subsubsection{UPB-based data augmentation}
Finally, we developed two data augmentation methods, one based on a global biased phase and the other on another frequency-based angle biased phase, and combined them with methods on the magnitude spectrum to augment input noisy speech.
The global biased phase-based data augmentation is consistent with the method in Sec. \ref{sec:hear_test} (\ref{eq:biased_stft_1}), so it is not detailed here. We focus on the linear biased phase-based data augmentation, which is similar to (\ref{eq:biased_stft_1}), but we replaced the $\theta$ with $\omega\tau'$, as shown in (\ref{eq:biased_stft_3}).
\begin{equation}
    \begin{aligned}
        X(t,\omega)&=\int^{\infty}_{-\infty}x(\tau)h(\tau-t)e^{-j(\omega\tau+\omega\tau')}d\tau\\
        &=\int^{\infty}_{-\infty}x(\tau)h(\tau-t)e^{-j\omega(\tau+\tau')}d\tau
    \end{aligned}
    \label{eq:biased_stft_3}
\end{equation}
Here, we set the value of $\tau'\in \frac{rand(0,2\pi)}{f_{s}}$, where $f_{s}$ means the sampling rate. From the relationship in the formula, we can see that this data augmentation method is equivalent to a time shift of $\tau'$ length in the time domain.

We also utilized a method of adding random real values on the magnitude spectrum, which follows the distribution of $\mathcal{N}(0,4\times10^{-6})$.

Note that the probability of using each of the above methods on each training audio sample is set to 50\%.

\vspace{-2.0mm}

\section{Experiments}
\vspace{-2.0mm}
\subsection{Dataset}
We conducted experiments on the VoiceBank-DEMAND dataset~\cite{demand}. The train set consists of 11,572 utterances (from 28 speakers) mixed with noise data with four signal-to-noise ratios (SNRs) (15, 10, 5, and 0 dB). The test set consists of 824 utterances (from two unseen speakers) mixed with unseen noise data with four SNRs (17.5, 12.5, 7.5, and 2.5 dB). All the utterances have been resampled to 16 kHz for a fair comparison.
\vspace{-2.0mm}

\subsection{Evaluation metrics}
Considering Table \ref{tab:pesq_ssnr_sisnr} in Sec. \ref{sec:hear_test}, which indicates the lack of correlation between time-domain metrics and perceived auditory quality, we abandoned the use of such metrics and only utilized the following four non-time-domain objective metrics to evaluate the performance of the proposed method and compare it with previous models. These included PESQ~\cite{pesq}, which assesses speech quality on a scale from --0.5 to 4.5, and the Short-time Objective Intelligibility (STOI)~\cite{stoi}, which evaluates speech intelligibility on a scale from 0 to 100\%. Two mean opinion score (MOS)-based measures were also considered, both of which operate on a scale from 1 to 5. These include CSIG~\cite{evaluation}, which predicts signal distortion, and COVL~\cite{evaluation}, which forecasts overall signal quality. For all these metrics, higher scores indicate a superior speech enhancement performance.

\subsection{Experimental configuration}
The training process involved segmenting the signals into 2-second chunks, with shorter signals being padded to meet this duration. The batch size was set to 32. All the models underwent training for a total of 180 epochs. To ensure optimal model selection, we saved the checkpoint after each epoch and recorded the best model on the validation set during the training progress. The optimizer we used was AdamW, with distinct initial learning rates for the generator ($4\times 10^{-3}$) and discriminator ($8\times 10^{-3}$). A StepLR scheduler with a gamma of 0.6 was utilized to adjust the learning rate every 30 epochs, facilitating a balanced learning pace. The $\lambda_{1\sim 7}$ weight parameters mentioned above are $[0.9,~0.1,~0.2,~0.05,~0.05,~0.05,~0.05]$. All the experiments were conducted using PyTorch (version 2.0.1) on 8 Nvidia RTX 3090Ti GPUs endowed with 24 GB of memory.

\vspace{-1.5mm}
\section{Results and ablation study}
\vspace{-1.5mm}
The results of the speech enhancement effects of the UPB-CMGAN model on the VoiceBank-DEMAND dataset, along with its ablation study results and comparisons with other models, are presented in Table \ref{tab:result}. The results indicate that replacing the original ri loss and time loss of CMGAN with UPB Loss led to an increase in PESQ, CSIG, and COVL. Furthermore, when we incorporated magnitude-based weights into the UPB Loss, these objective evaluation metrics showed additional improvement. A similar trend was observed when the phase derivative input was added to the original discriminator, which was solely based on magnitude. With an informal listening of generated samples, we observed a slight hum noise recognized in some worse-case samples. However, the objective metrics used to evaluate the performance of the methods on VoiceBank-DEMAND indicated that leveraging UPB-based data augmentation had a synergistic effect that led to the aforementioned improvements. Overall, our model achieved the PESQ of 3.55, CSIG of 4.78, and COVL of 4.28, which are SoTA results.

\vspace{-1.5mm}
\section{Conclusion}
\vspace{-1.5mm}
In this paper, we propose four mutually compatible improvements for the estimation of phase spectra based on the CMGAN framework. We abandon the traditional approach of precise phase reconstruction and instead estimate phases with unrestricted global-phase bias to reduce the training burden of the model.
We conducted thorough ablation experiments on the proposed UPB-CMGAN model and compared it with various existing models. The results indicate that all four proposed methods are reasonable and effective. Furthermore, when these methods were implemented collaboratively, the UPB-CMGAN achieved a SoTA performance on the VoiceBank-DEMAND dataset without incurring additional computational costs.

\vfill\pagebreak

\footnotesize

\vfill\pagebreak

\onecolumn

\centerline{\Huge \textbf{Appendix}}

\vspace{3.mm}

\normalsize This paper was accepted and presented at ICASSP 2024. After presenting it at ICASSP 2024, we recognized that a data augmentation scheme~\cite{phaseaug} similar to the "frequency-based angle biased phase" part~(\ref{eq:biased_stft_3}) of our method in Section 3.2.4 has been proposed in ICASSP 2023. Therefore, we added a reference to the paper here.

\end{document}